\documentclass[prc,aps,eqsecnum,amssymb,floatfix]{revtex4}
\usepackage{bm}
\usepackage{graphicx}
\newcommand{\bfx}{{\bm x}}
\newcommand{\bfr}{{\bm r}}
\newcommand{\bfk}{{\bm k}}
\newcommand{\bfp}{{\bm p}}

\begin{document}
\title{Variational Calculation on $A$=3 and 4 Nuclei with Non-Local Potentials}
\author{M. Viviani$^{(a,b)}$, L. E. Marcucci$^{(b,a)}$, S. Rosati$^{(b,a)}$,
A. Kievsky$^{(a,b)}$, and L. Girlanda$^{(b)}$}
\affiliation{
(a) INFN, Sezione di Pisa, I-56100 Pisa, Italy
\\
(b) Department of Physics ``E. Fermi'', University of Pisa, 
I-56127 Pisa, Italy}
\date{\today}

\begin{abstract}
The application of the hyperspherical harmonic approach
to the case of non-local two-body potentials is described. Given 
the properties of the hyperspherical harmonic functions, 
there are no difficulties in considering the approach in 
both coordinate and momentum space.   
The binding energies and other
ground state properties of $A=3$ and 4 nuclei are calculated 
using the CD Bonn 2000 and N3LO
two-body potentials. The results 
are shown to be in excellent agreement with
corresponding ones obtained by other accurate techniques.
\end{abstract}


\maketitle

\section{Introduction}
\label{sec:intro}

A number of ``realistic''
nucleon-nucleon ($NN$) potentials 
have been determined in recent years~\cite{AV18,Nijm,Mac01}, that 
reproduce the deuteron binding energy and fit a large
set of $NN$ scattering data below the pion-production threshold
with a $\chi^2$/datum $\simeq$ 1. The high accuracy 
achieved in reproducing $NN$
observables has spurred renewed interest in testing these 
potential models in systems with
$A\ge 3$. Several methods have been developed to accurately solve 
the Schr\"odinger equation for these systems, and meaningful
comparisons with precise experimental data, 
for example for the $N-d$ reaction, are nowadays possible~\cite{GWHKG96,Car98}.
Among these methods, of particular importance are the Faddeev-Yakubovsky
(FY) equations approach~\cite{Glo83,Nog00,Nea02,Cie99,Del03,Del05}, 
the quantum Monte
Carlo methods~\cite{Pud95,Pud97}, and the techniques based 
on the expansion of the nuclear wave function 
on an appropriate basis, like the hyperspherical-harmonics 
(HH)~\cite{Kie93,Kie94,Kie97a,Viv05,Bar00}, 
the harmonic oscillator~\cite{Nav99,Nav04}, or 
the gaussian basis~\cite{Kam89,Suz98}. Some of these 
methods are variational
while others are not, and they are more or less advantageous depending on
the problem at hand. All of them provide very accurate results  
for bound states~\cite{Kam01}. On the other hand, the application to
scattering states presents a number of difficulties that are 
specific to each of these methods. 
Of course, scattering wave functions are an essential input to the 
calculation of scattering observables as well as 
of cross sections for electroweak reactions, such as, 
for example, the low energy radiative and weak capture processes 
relevant in nuclear astrophysics. 
The FY approach has been developed both in coordinate and
momentum space and, in general, gives very accurate
results. However, this approach becomes problematic when considering 
scattering processes with charged particles at low energies, 
although recent improvements have been made
in Ref.~\cite{Del05} to reduce some of these problems. 
In this respect, variational methods 
do not encounter any difficulty, since the electromagnetic 
interaction between the particles can be taken 
into account in configuration space~\cite{Kie04}.  
For example, with the HH technique 
$p-d$ elastic scattering observables have been calculated with
the same degree of accuracy as corresponding $n-d$
ones~\cite{Kie01a,Kie01b,Del05b}.

The failure of theory based the ``realistic'' $NN$ 
models in predicting successfully $A$=3 and 4 bound and scattering state 
properties provides evidence for the need of including 
a three nucleon interaction (TNI)~\cite{Wea01,Erm05}.
However, the current understanding of the TNI is still in an early
stage, and discrepancies between theory and experiments 
are still observed, as in the case of the
$N-d$~\cite{Erm05,KH87,WGC88,Kie96} and  $p-{}^3$He~\cite{KRV01,PHH01} $A_y$
polarization observable (the so-called ``$A_y$-puzzle'')
It is not clear if these 
discrepancies can be solved by using more sophisticated models of 
TNIs, or if the problem still resides in the $NN$ interaction
(in particular, in $P$ waves)~\cite{HF98,TT99,EMW02}. 

More recently, new potentials have been derived using chiral perturbation  
theory at increasing order~\cite{EGM00,EM03}. Also many-body forces can be
derived on the same footing~\cite{FHK99,K94,Eea02,FPK05,EMP05}. 
Furthermore, there are other models which are constructed 
up to a certain cutoff momentum $\Lambda$
({\it low-q} potentials~\cite{BKS03,Cor05}). 
All these new potentials are defined primarily in
momentum space and are non-local. Therefore, 
accurate techniques which can solve the corresponding Schr\"odinger
equation for $A$=3 and 4 are important. 
At present, only the FY~\cite{Nea02,Del03} and the
No Core Shell Model~\cite{Nav04} methods have been employed for this 
task. In this paper, we will show that the HH method too can be successfully
applied to treat this kind of potentials. 

In the last few years, considerable effort has been devoted by the
authors of the present paper to the development and 
application of the HH technique to 
study bound and scattering states of three or four 
nucleons with realistic local $NN$ 
potentials~\cite{Kie93,Kie94,Kie97a,Viv95,Kie95,Viv98}.
A version of the method, that 
has been rather extensively exploited in these calculations, 
includes an appropriate correlation factor in the HH basis functions, 
so as to take into account 
the strong short-range repulsion of the $NN$ interaction
and, therefore, improve the rate of convergence of the expansion.  
This approach is known as the correlated-hyperspherical-harmonics
(CHH) method~\cite{Kie93,Kie94,Viv95}.  
Up until now, the calculations have been done 
only for local nuclear interactions, in particular 
the Argonne $v_{18}$ two-nucleon (AV18)~\cite{AV18} together with the
Urbana IX (UIX) three-nucleon~\cite{Pud95} interaction.

The HH method, without correlation
factors, has been also employed for the calculation of
trinucleon~\cite{Kie97a} and $\alpha$ particle~\cite{Viv05} bound states
and four-nucleon scattering processes~\cite{Laz05}. 
Such a method appears especially convenient 
in the case of non-local potentials. The object of the present
paper is the calculation of the trinucleon and $\alpha$ particle bound 
states with the uncorrelated HH expansion using 
two recent, realistic non-local two-nucleon interactions,  
namely the CD Bonn 2000~\cite{Mac01} and the N3LO~\cite{EM03} potentials.
This kind of expansion, in fact, can be performed equally well in
coordinate or momentum space~\cite{Ros01}. It is thus possible 
to treat in the corresponding space
the part of interaction given in coordinate or momentum space.
With two-body local potentials the matrix elements of the interactions can be
obtained via two dimensional integrals. The non-locality of the 
interaction merely requires a three-dimensional integration, which 
can be performed using standard numerical methods. Obviously, there are no
difficulties in including the local Coulomb interaction. The same 
is valid for the most commonly used TNIs, which are local 
interactions.
However, the TNI has not been considered 
in the calculations reported here. 
As a final remark, it should be noted that 
the present work is only the ``starting point'' for the 
implementation of the HH method 
in the study of $A$=3 and 4 scattering states
with non-local interactions.

The paper is organized as follows. The details of the formalism
are presented in the next section. In Sec.~\ref{sec:res}, the 
results for the binding energies and ground state properties 
of systems with $A$=3 and 4 
are presented and compared with those 
obtained with other different techniques.
A few final remarks and conclusions are given 
in Sec.~\ref{sec:con}.

\section{Formalism}
\label{sec:cal}

In general, an $A$-body bound state can be written as
\begin{equation}
  |\Psi(1,2,\cdots,A)\rangle=\sum_{\mu }\, c_\mu\, 
  |\Psi_\mu \rangle \ ,
  \label{eq:psi2}
\end{equation}
where $|\Psi_\mu\rangle$ are a suitable complete set of states, 
and $\mu$ is an index denoting the set of 
quantum numbers necessary to completely determine 
the basis elements.
The coefficients of the expansion can be calculated using the 
Rayleigh-Ritz variational principle, which states that
\begin{equation}
  \langle\delta_c \Psi(1,2,\cdots,A)\,|\,H-E\,|\Psi(1,2,\cdots,A)\rangle
   =0 \ ,
   \label{eq:rrvar}
\end{equation}
where $\delta_c \Psi(1,2,\cdots,A)$ indicates the variation of 
$\Psi(1,2,\cdots,A)$ for arbitrary infinitesimal 
changes of the linear coefficients $c_\mu$. 
The problem of determining $c_\mu$ and the energy $E$ 
is then reduced to a generalized eigenvalue problem, 
\begin{equation}
  \sum_{ \mu'}\,\langle\Psi_\mu\,|\,H-E\,|\, \Psi_{\mu'}\,\rangle \,c_{\mu'}=0
  \ .
  \label{eq:gepb}
\end{equation}
The main difficulty of the method is to compute the 
matrix elements of the Hamiltonian $H$ with respect to the basis states
$|\Psi_\mu\rangle$. Usually $H$ is given as a sum of terms (kinetic energy,
two-body potential, etc.). The calculation of the matrix elements of
some parts of $H$ can be more conveniently performed in coordinate 
space, while for other parts it could be easier to work in momentum
space. Therefore, it is important that the basis states 
$|\Psi_\mu\rangle$ have simple expressions in both spaces. The
HH functions do have such a property, as will be shown below.

Let us first consider the expression of the HH functions in coordinate
space. The internal dynamics of a system of $A$ identical nucleons of mass
$m$ is conveniently described in terms of a
set of $N=A-1$ Jacobi vectors $\bfx_{1p},\ldots,\bfx_{Np}$,
constructed from a given particle permutation denoted with $p$, which 
specifies the particle order 
$i,j,k,\ldots$. In particular,
$\bfx_{Np}=(\bfr_j-\bfr_i)/\sqrt{2}$ ($\bfr_i$ is the position of the
$i$-th particle, etc.), and $p=1$ is chosen to correspond to the 
particle order $1,2,3,\ldots$. The hyperradial coordinates are defined to be 
\begin{eqnarray}
  \rho&=&\sqrt{\sum_{i=1}^N\bfx_{ip}^2} \ , \label{eq:rho} \\
  \Omega^{(\rho)}_p&=&[{\hat{\bfx}}_{1p},{\hat{\bfx}}_{2p},\cdots,
                 {\hat{\bfx}}_{Np};
  \phi_{2p},\cdots,\phi_{Np}] \ , 
  \label{eq:omegar}
\end{eqnarray}
where $\rho$ is the so-called hyperradius, $\Omega_p^{(\rho)}$ a
set of angular-hyperangular coordinates
and the suffix $(\rho)$ recalls the use of the coordinate space.  
Note that $\rho$ does not depend on
the particle permutation used to construct the Jacobi vectors. The angles 
$\phi_{2p},\cdots,\phi_{Np}$ are the hyperangles, 
defined as~\cite{Fab83}
\begin{equation}
  \tan{\phi_{ip}}=\frac{1}{x_{ip}}\,\sqrt{\sum_{j=1}^{i-1}\,x_{jp}^2}\ ,
  \qquad i=2,\ldots,N  \ .
  \label{eq:hypera}
\end{equation}
In terms of these variables, chosen any particle permutation $p$, 
the kinetic energy operator becomes
\begin{equation}
  T=-\frac{\hbar^2}{2m}\sum_{i=1,N}\nabla^2_{\bfx_{ip}} =
    -\frac{\hbar^2}{2m}\left( \frac{ \partial^2}{\partial\rho^2} +
     \frac{3N-1}{\rho}\frac{\partial}{\partial\rho}-
     \frac{\Lambda^2(\Omega^{(\rho)}_p)}{\rho^2}
     \right) \ .
  \label{eq:kinetic}
\end{equation}
The HH functions $Y_{[G]}(\Omega^{(\rho)}_p)$ are the eigenfunctions
of the operator $\Lambda^2(\Omega^{(\rho)}_p)$, 
and their explicit expression for a generic $A$-nucleon system can be 
found, for example, in Refs.~\cite{Fab83,Ros01}. 
Here we consider in some details
only the $A$=3 case. The Jacobi coordinates for $A$=3 nucleons 
are defined as
\begin{eqnarray}
  \bfx_{2p}&=&\frac{1}{\sqrt{2}}(\bfr_j-\bfr_i) \ , \nonumber \\
  \bfx_{1p}&=&\sqrt{\frac{2}{3}}(\bfr_k-\frac{1}{2}(\bfr_i+\bfr_j)) \ , 
  \label{eq:jacc3}
\end{eqnarray}
where $p=1$ corresponds to the order 1,2,3.
The HH function $Y^{LL_z}_{ [G] }(\Omega^{(\rho)}_p)$, with a definite value
of the total orbital angular momentum $L,L_z$, can be written as~\cite{Kie97a}:
\begin{equation}
  Y^{LL_z}_{ [G] }(\Omega^{(\rho)}_p) =
  \biggl[Y_{\ell_2}({\hat{\bfx}}_{2p}) \otimes
   Y_{\ell_1}({\hat{\bfx}}_{1p}) \biggr]_{LL_z}
  N_{[G] }\,(\cos\phi_p)^{\ell_2}(\sin\phi_p)^{\ell_1}\,
  P_{n_2}^{\ell_1+\frac{1}{2},\ell_2+\frac{1}{2}}(\cos 2\phi_p) \ .
\label{eq:hh3}
\end{equation}
Note that in the $A=3$ system only the hyperangle $\phi_{2p}$ is 
present (indicated with $\phi_p$). In Eq.~(\ref{eq:hh3}), 
$Y_{\ell_1}({\hat{\bfx}}_{1p})$ and $Y_{\ell_2}({\hat{\bfx}}_{2p})$ are 
spherical harmonics on the two internal Jacobi coordinates 
$\bfx_{1p}$ and $\bfx_{2p}$, $N_{[G]}$ is a normalization factor and 
$P_{n_2}^{l_1+\frac{1}{2},l_2+\frac{1}{2}}(\cos 2\phi_p)$ is a Jacobi 
polynomial, $n_2$ being the degree of the polynomial.
The grand angular quantum number $G$ is 
defined as $G=2n_2+\ell_1+\ell_2$.
The notation $[G]$ stands for $[\ell_1,\ell_2;n_2]$. 

Moreover, we consider the anti-symmetrized functions 
${\cal Y}_{ \{G\}}(\Omega^{(\rho)})$ given by the product of 
an HH and a spin-isospin function.
The antisymmetry is obtained by writing 
${\cal Y}_{ \{G\}}(\Omega^{(\rho)})$ as a sum of terms constructed 
by starting from all the possible even permutations $p$ of the particles,
assuming antisymmetry in the interacting pair. 
For $A=3$ they are explicitly given by
\begin{equation}
  {\cal Y}_{ \{G\} }(\Omega^{(\rho)})= \sum_{p}^{even}\bigg[ 
  Y^{LL_z}_{ [G] }(\Omega^{(\rho)}_p) 
  \otimes [S_2\otimes \frac{1}{2}]_{S S_z} \bigg]_{J J_z}\, 
  [T_2\otimes\frac{1}{2}]_{T T_z}   \ , \label{eq:hha3}
\end{equation}
where the spins (isospins) of particle $i$ and $j$ are coupled to 
$S_2$ ($T_2$), which is itself coupled to the spin (isospin) 
of the third particle to give 
the state with total spin $S$ (isospin $T,T_z$).
The total orbital angular momentum $L$ and the total 
spin $S$ are coupled to the total angular momentum 
$J,J_z$.
Each set of quantum numbers $\{\ell_1,\ell_2,L,S_2,T_2,S,T\}$ 
is called ``channel'', and here the notation $\{ G\}$ stands 
for 
\begin{equation}
  \{ G\}\equiv\{\ell_1,\ell_2,L,S_2,T_2,S,T;n_2\} \ .
  \label{eq:G}
\end{equation}
The requirement of the antisymmetry of ${\cal Y}_{ \{G\} }(\Omega^{(\rho)})$
constraints the possible choices
of the set $\{ G\}$ to those for which
\begin{equation}
   (-)^{\ell_2+S_2+T_2}=-1\ .\label{eq:lst}
\end{equation}
The corresponding expressions for the $A=4$ system 
can be found in Ref.~\cite{Viv05}.

In this paper, we will consider modern two-body potential models 
which act on specific spin and angular momentum states of the 
two-body system. Due to the presence of the sum over the permutations in
the expression for ${\cal Y}_{ \{G\} }$, 
a given particle pair
is not in a definite angular and spin state. However,
the HH functions with the grand angular quantum number $G$ constructed in
terms of a given set of Jacobi vectors
$\bfx_{1p},\cdots,\bfx_{Np}$, defined starting from the
order $i,j,k,\ldots$ of particles, can be always expressed in terms of
the HH functions constructed, for instance, in terms of
$\bfx_{1 (p=1)},\cdots,\bfx_{N (p=1)}$.  
For example, when $A=3$, the following relation holds
\begin{equation}
  Y^{LL_z}_{[\ell_1,\ell_2;n_2]}(\Omega^{(\rho)}_p) =
   \sum_{\ell_1',\ell_2',n_2'}
    a^{(p),L}_{\ell_1,\ell_2,n_2;\,\ell_1',\ell_2',n_2'}
    Y^{LL_z}_{[\ell_1',\ell_2';n_2']}(\Omega^{(\rho)}_{(p=1)})\ ,
   \label{eq:rr3}
\end{equation}
where the HH functions are defined in Eq.~(\ref{eq:hh3}) and the sum is
restricted to the values $\ell_1'$, $\ell_2'$, and $n_2'$
such that $\ell_1'+\ell_2'+2n_2'=G$. 
The coefficients $ a^{(p)}_{\ell_1,\ell_2,n_2;\,\ell_1',\ell_2',n_2'}$
relating the two sets of HH functions are known as the Raynal-Revai
coefficients~\cite{RR70}. There  exist several procedures to compute these
coefficients for $A=3$ and $4$
systems~\cite{RR70,BM90,DLWD95,NK94,E95,V98}. For $A=3$,   
they can be computed rather easily using the orthonormality property of the HH
functions, namely 
\begin{equation}
  a^{(p),L}_{\ell_1,\ell_2,n_2;\,\ell_1',\ell_2',n_2'}=
  \int d\Omega^{(\rho)}_{(p=1)} \left(
   Y^{LL_z}_{[\ell_1,\ell_2;n_2]}(\Omega^{(\rho)}_{(p=1)})\right)^*
   Y^{LL_z}_{[\ell_1',\ell_2';n_2']}(\Omega^{(\rho)}_p)\ .
   \label{eq:rr3b}
\end{equation}
For $A=4$, we have used the procedure devised in Ref.~\cite{V98}, 
where the corresponding coefficients are obtained by using a 
set of recurrence relations.
Also the spin-isospin states can be recoupled to obtain states where the
spin/isospin are coupled in a given order of the particles. 
The result is that the antisymmetric functions 
${\cal Y}_{ \{G\} }$ can be expressed as a
superposition of functions constructed in terms of a given order of particles
$i,j,k,\ldots$, each one having the pair $i$,$j$ in a definite spin and
angular momentum state. When the two-body potential acts on the pair of
particles $i$,$j$, the effect of the projection is easily taken into
account. 

We now consider the expansion states $|\Psi_\mu\rangle$ of 
Eq.~(\ref{eq:psi2}). In coordinate space, they have been chosen 
to be given by 
\begin{equation}
  \langle\,\bfx_1,\ldots,\bfx_N\,|\,\Psi_\mu\,\rangle
  = f_l(\rho) {\cal Y}_{ \{G\} }(\Omega^{(\rho)})\ ,
  \label{eq:rexp}
\end{equation}
where $\mu$ stands for the set of quantum numbers $(\{G\},l)$,
and $f_l(\rho)$ for $l=1,\ldots$ is a complete set of hyperradial 
functions. Note that the index $p$ has been suppressed, since 
$|\Psi_\mu\rangle$ is at this stage independent on the permutation.
The corresponding states in momentum space can be obtained as follows.
Let $\hbar\bfk_{1p},\cdots,\hbar\bfk_{Np}$ be the conjugate Jacobi momenta 
of the Jacobi vectors. Explicitly, in the $A$=3 case, these momenta
are
\begin{eqnarray}
  \hbar\bfk_{2p}&=&\frac{1}{\sqrt{2}}(\bfp_j-\bfp_i) \ , \nonumber \\
  \hbar\bfk_{1p}&=&\sqrt{\frac{2}{3}}(\bfp_k-\frac{1}{2}(\bfp_i+\bfp_j)) \ , 
  \label{eq:jacm3}
\end{eqnarray}
$\bfp_i$ being the momentum of the $i$-th particle. 
The following relation 
then may be shown to hold~\cite{Fab83,Ros01}:
\begin{eqnarray}
  \lefteqn{\langle\,\bfx_{1p},\ldots,\bfx_{Np}\,|\,
    \bfk_{1p},\ldots,\bfk_{Np}\,\rangle
   =}\qquad\qquad
   \qquad &&  \nonumber \\
  &=& \frac{1}{(2\pi)^{3N/2}}\,  
    {\rm exp}\bigg( i\sum_{j=1}^N \bfk_{jp}\cdot\bfx_{jp} \bigg)= \nonumber \\
  &=& \frac{1}{ (Q\rho)^{3N/2-1}}
  \sum_{ \{G\} } i^G\,Y_{ [G] }(\Omega^{(\rho)}_p)
                      Y^*_{ [G] }(\Omega^{(Q)}_p)
  J_{{\cal L}+1/2}(Q\rho) \ ,
\label{eq:ekr}
\end{eqnarray}
where ${\cal L}=G+(3N-3)/2$, $J_{{\cal L}+1/2}(Q\rho)$ is a Bessel 
function, $\Omega^{(\rho)}_p$ is given in Eq.~(\ref{eq:omegar}), and 
\begin{equation}
  Q=\sqrt{\sum_{i=1}^N\bfk_{ip}^2} \ , 
  \label{eq:hyperm}
\end{equation}
is the hypermomentum, which plays the role in momentum space 
corresponding to $\rho$ in coordinate space. Also $Q$ does  not depend on the
particular permutation used to construct the vectors
$\hbar\bfk_{1p},\cdots,\hbar\bfk_{Np}$. The momentum-space 
angular-hyperangular variables  are defined as
\begin{equation}
  \Omega^{(Q)}_p=[{\hat{\bfk}}_{1p},{\hat{\bfk}}_{2p},\cdots,
                  {\hat{\bfk}}_{Np};
  \varphi_{2p},\cdots,\varphi_{Np}] \ , 
  \label{eq:omegaq}
\end{equation}
and
\begin{equation}
  \tan{\varphi_{ip}}=\frac{1}{k_{ip}}\,\sqrt{\sum_{j=1}^{i-1}\,k_{jp}^2}\ ,
   \qquad i=2,\dots,N  \ .
  \label{eq:hyperaq}
\end{equation}
Then, the momentum space version of the wave function 
given in Eq.~(\ref{eq:rexp}) is
\begin{equation}
  \langle\,\bfk_{1},\ldots,\bfk_{N}\,|\,\Psi_\mu\,\rangle=
  \int d\bfx_{1p}\cdots d\bfx_{Np}\,
  \frac{{\rm e}^{-i\sum_{j=1}^N \bfk_{jp}\cdot\bfx_{jp}}}{(2\pi)^{3N/2}}
   f_l(\rho) {\cal Y}_{ \{G\} }(\Omega^{(\rho)})\ .
   \label{eq:ft}
\end{equation}
Note that also 
$\langle\,\bfk_{1},\ldots,\bfk_{N}\,|\,\Psi_\mu\,\rangle$ cannot 
depend on the permutation index $p$. In fact, 
using the expansion of Eq.~(\ref{eq:ekr}) and the orthogonality property of
the HH functions, it follows that 
\begin{equation}
  \langle\,\bfk_1,\ldots,\bfk_N\,|\,\Psi_\mu\,\rangle=
   g_{ G,l }(Q) {\cal Y}_{ \{G\} }(\Omega^{(Q)}) \ , 
  \label{eq:qexp}
\end{equation}
where ${\cal Y}_{ \{G\} }(\Omega^{(Q)})$ is the same as 
${\cal Y}_{ \{G\} }(\Omega^{(\rho)})$ with $\bfx_{ip}\rightarrow\bfk_{ip}$, 
and 
\begin{equation}
   g_{G,l}(Q)=(-i)^G\,\int_0^\infty d\rho\,
   \frac{\rho^{3N-1}}{(Q\rho)^{3N/2-1}}\,
   J_{{\cal L}+1/2}(Q\rho)\, f_{l}(\rho) \ .
\label{eq:vg}
\end{equation}
In the present work, we have adopted two choices for the  functions 
$f_{ l }(\rho)$, for which the Fourier transform given in
Eq.~(\ref{eq:vg}) can be obtained (almost) analytically. The two choices are
illustrated below.

\begin{enumerate}

\item Exponential functions of the type
\begin{equation}
  f_l(\rho)= {\rm e}^{-\alpha_l\rho} \ ,
  \label{eq:flexp}
\end{equation}
where $\alpha_1=\beta$ and $\alpha_l=\beta+\gamma^{l-1}$ for
$l=2,\cdots M$,  $\beta$ and $\gamma$ being two non-linear 
parameters, which need to be optimized depending on the used nuclear 
potential model. As an example, when $A=3$, $\beta=0.78$ fm$^{-1}$ and 
$\gamma=1.2$ fm$^{-1}$ for the CD Bonn 2000 potential. Other 
choices of $\alpha_l$ have been made, for instance 
$\alpha_l=\beta+\gamma(l-1)$ or $\alpha_l=\beta\gamma^{l-1}$, 
but no significant differences have been found in the 
final results, once the non-linear parameters $\beta$ 
and $\gamma$ are optimized and convergence on $M$ has been reached. 
The advantage of using the 
exponential basis~(\ref{eq:flexp}) is that the corresponding functions 
$g_{G,l}(Q)$ have an easy expression, and are given by:
\begin{equation}
   g_{G,l}(Q)=(-i)^G\, \frac{\Gamma(D+G)}{\alpha_l^D}
          \frac{u^D}{(1-u^2)^{\frac{D}{4}-\frac{1}{2}}}
          P_{D/2}^{1-G-D/2}(u)\ , \qquad
    u=\sqrt{\frac{1}{1+Q^2/\alpha_l^2}}\ ,
   \label{eq:glexpA}
\end{equation}
where $D=3N$ and $P_n^m$ is an associated Legendre function. For $A=3$ this
expression can be written in terms of analytical functions, namely
\begin{equation}
   g_{G,l}(Q)=-\frac{(-i)^G}{Q^{G+4}}
   \; \frac{\sqrt{5!}}{(2\alpha_l)^3} \,\, \frac{d^3}{d\alpha_l^3}
   \bigg[ \frac{ (\sqrt{\alpha_l^2+Q^2}-\alpha_l)^{G+2} }
  {\sqrt{\alpha_l^2+Q^2}}\bigg] \ . 
\label{eq:glexp}
\end{equation}

\item  Another useful form for $f_l(\rho)$, adopted 
for example also in Ref.~\cite{Kie97b}, is
\begin{equation}
 f_l(\rho)=\gamma^{D/2} \sqrt{\frac{l!}{(l+D-1)!}}\,\,\, 
 L^{(D-1)}_l(\gamma\rho)\,\,{\rm e}^{-\frac{\gamma}{2}\rho} \ ,
 \label{eq:fllag}
\end{equation}
where $L^{(D-1)}_l(\gamma\rho)$ are Laguerre polynomials. 
Here, there is only one non-linear parameter, $\gamma$,  
to be variationally optimized. In particular, 
$\gamma$ can be chosen in the interval 
3.5--4.5 fm$^{-1}$ for the CD Bonn 2000 potential and 
6--8 fm$^{-1}$ for the N3LO potential, for both $A=3$ and 4. 
The corresponding functions $g_{G,l}(Q)$ are less trivial to calculate, 
and are given by:
\begin{equation}
  g_{G,l}(Q)=\frac{(-i)^G}{\gamma^{D/2}}
  \sqrt{\frac{l!}{(l+D-1)!}}\, \sum_{k=0}^l b^l_k\,2^{k+D}\,\,
  \Gamma(G+k+D)\,\,\frac{u^{k+D}}{(1-u^2)^{\frac{D}{4}-\frac{1}{2}}}\,\,
  P^{1-G-D/2}_{k+D/2}(u) \ ,
\label{eq:gllag}
\end{equation}
where $u=\frac{1}{\sqrt{1+(2Q/\gamma)^2}}$ and $b^l_k$ are given by
\begin{equation}
   b^l_k=\frac{(-1)^k}{k!}\, \left( \begin{array}{c}
                                l+D-1 \\ l-k 
                               \end{array} \right)
  \ ,
  \label{eq:ak}
\end{equation}
so that $L^{(D-1)}_l(x)=\sum_{k=0}^l b^l_k\,x^k$~\cite{Abr72}.
Studies on the convergence on $M$ for the $A$=3 and 4 calculations are
presented in the Sec.~\ref{sec:res}.

\end{enumerate}

In summary, the variational state is given by
\begin{equation}
  |\,\Psi(1,\ldots,A)\,\rangle=
  \sum_{ \{G\} } \sum_{l=1}^M c_{ \{G\},l }  
   |\, \Psi_{ \{G\},l } \,\rangle \ ,  
  \label{eq:wafu1}
\end{equation}
where, in momentum space the expansion states are
\begin{equation}
  \langle\,\bfk_1,\ldots,\bfk_N\,| \, \Psi_{ \{G\},l } \,\rangle =
   g_{ G,l }(Q) {\cal Y}_{ \{G\} }(\Omega^{(Q)}) \ ,
  \label{eq:wafu2}
\end{equation}
and in coordinate space are
\begin{equation}
  \langle\,\bfx_1,\ldots,\bfx_N\,| \, \Psi_{ \{G\},l } \,\rangle =
   f_{ l }(\rho) {\cal Y}_{ \{G\} }(\Omega^{(\rho)}) \ .
  \label{eq:wafu3}
\end{equation}
These two expressions can be used to evaluate the matrix
elements in Eq.~(\ref{eq:gepb}). In particular, 
the normalization ($N$) and kinetic
energy ($T$) operator matrix elements can be computed both in coordinate 
and in momentum space. Explicitly:
\begin{eqnarray}
  N_{\{G'\},l';\{G\},l}&=& \int\,d\rho\,\rho^{3N-1}\,f_{l'}(\rho)\,f_{l}(\rho)
  \int\,d\Omega^{(\rho)}\,{\cal Y}^*_{ \{G'\} }(\Omega^{(\rho)})\,
  {\cal Y}_{ \{G\} }(\Omega^{(\rho)}) \nonumber \\
  &=&\int\,dQ\,Q^{3N-1}\,g_{G',l'}(Q)\,g_{G,l}(Q)
  \int\,d\Omega^{(Q)}\,{\cal Y}^*_{ \{G'\} }(\Omega^{(Q)})\,
  {\cal Y}_{ \{G\} }(\Omega^{(Q)}) \ ,
  \label{eq:norm} \\
  T_{\{G'\},l';\{G\},l}&=& -\frac{\hbar^2}{2m}
  \int\,d\rho\,\rho^{3N-1}\,f_{l'}(\rho)\,
  \bigg[ \frac{\partial^2}{\partial\rho^2}
      +\frac{3N-1}{\rho}\frac{\partial}{\partial\rho}
      -\frac{G(G+3N-2)}{\rho^2} \bigg]\,f_{l}(\rho)\nonumber \\
  &&\times \int\,d\Omega^{(\rho)}\,{\cal Y}^*_{ \{G'\} }(\Omega^{(\rho)})\,
  {\cal Y}_{ \{G\} }(\Omega^{(\rho)}) \nonumber \\
  &=& \frac{\hbar^2}{2m}\,\int\,dQ\,Q^{3N+1}\,g_{G',l'}(Q)\,g_{G,l}(Q)
  \nonumber \\
&&\times  \int\,d\Omega^{(Q)}\,{\cal Y}^*_{ \{G'\} }(\Omega^{(Q)})\,
  {\cal Y}_{ \{G\} }(\Omega^{(Q)}) \ ,
  \label{eq:kin}
\end{eqnarray}
where it has been used the fact that the HH functions are 
eigenfunctions of the operator $\Lambda^2(\Omega^{(\rho)}_p)$ defined in
Eq.~(\ref{eq:kinetic}) corresponding to the eigenvalues $G(G+3N-2)$. 
In Eqs.~(\ref{eq:norm}) and~(\ref{eq:kin}), as well as in the rest 
of the present work, the permutation index $p$ is omitted, 
and the integration variables corresponding to $p=1$ 
are used (i.e.
$d\Omega^{(\rho/Q)}\equiv d\Omega^{(\rho/Q)}_{(p=1)}$).

The calculation of the two-body potential energy matrix elements is more
conveniently 
performed in either coordinate or momentum space, depending on the particular
potential model of interest. 
First of all, due to the antisymmetry of the wave function, the 
following relation holds
\begin{equation}
V_{\{G'\},l';\{G\},l}\equiv
  \langle\Psi_{ \{G'\}, l' }|V|\Psi_{  \{G\}, l}\rangle = \frac{A(A-1)}{2}
  \langle\Psi_{ \{G'\}, l' }|v(1,2)|\Psi_{  \{G\}, l}\rangle\ ,
  \label{eq:v12}
\end{equation}
where $v(1,2)$ acts on the particle pair $1$,$2$ and can be written as
\begin{equation}
v(1,2)=V(\bfx'_N;\bfx_N) \ ,
\label{eq:v12cs}
\end{equation}
in coordinate space, and
\begin{equation}
v(1,2)=\widetilde V(\bfk'_N;\bfk_N) \ ,
\label{eq:v12ms}
\end{equation}
in momentum space (clearly $\widetilde V$ and $V$ are related by a Fourier
transform). Then,  
\begin{eqnarray}
  V_{\{G'\},l';\{G\},l} &=& \frac{A(A-1)}{2} \int\,d\bfx_1\cdots d\bfx_N
  \int\,d\bfx'_N\,\, 
  f_{l'}(\rho')\,{\cal Y}^*_{ \{G'\} }(\Omega^{(\rho')}) \nonumber \\
 && \qquad\times V(\bfx'_N;\bfx_N)
    f_{l}(\rho)\,{\cal Y}_{ \{G\} }(\Omega^{(\rho)}) \label{eq:potr} 
\end{eqnarray}
or
\begin{eqnarray}
  V_{\{G'\},l';\{G\},l} &=&  \frac{A(A-1)}{2}\int\,d\bfk_1\cdots d\bfk_N 
\int\,d\bfk'_N\,\, 
  g_{G',l'}(Q')\,{\cal Y}^*_{ \{G'\} }(\Omega^{(Q')}) \nonumber \\
 &&\qquad\times \widetilde V(\bfk'_N;\bfk_N)
   g_{G,l}(Q)\,{\cal Y}_{ \{G\} }(\Omega^{(Q)}) \ ,
  \label{eq:potq}
\end{eqnarray}
where $\rho',\Omega^{(\rho')}$ 
are the hyperradial coordinates
associated to the Jacobi vectors $\bfx_1,\cdots,\bfx_{(N-1)},\bfx'_N$, etc.
When the potential energy operator
is a local operator in coordinate space, namely
\begin{equation}
   V(\bfx'_N;\bfx_N)\rightarrow
   V_{loc}(\bfx_N) \delta(\bfx'_N-\bfx_N)\ ,\label{eq:localr}
\end{equation}
it is more convenient to calculate the corresponding 
matrix elements using Eq.(\ref{eq:potr}), which
simplifies to
\begin{equation}
  V_{\{G'\},l';\{G\},l}=
  \int\,d\bfx_1\cdots d\bfx_N  
  f_{l'}(\rho)\,{\cal Y}^*_{ \{G'\} }(\Omega^{(\rho)})\,
  V_{loc}(\bfx_N)
    f_{l}(\rho)\,{\cal Y}_{ \{G\} }(\Omega^{(\rho)})\ . \label{eq:potrloc}
\end{equation}
Examples of two-nucleon potential models of this form are the Argonne
$v_{18}$~\cite{AV18} or the Nijmegen II~\cite{Nijm} potentials. 
Note that previous applications of the HH method were limited 
to these cases in Refs.~\cite{Kie97a,Viv05}.
On the other hand, for the CD Bonn 2000 or N3LO potentials,
which are non-local operators in momentum space, the use of the
momentum-space expression of $ V_{\{G'\},l';\{G\},l}$ is more
convenient. In general, the potential energy operator is given as
a sum of different local and/or non-local terms. 
The computation of the matrix elements of
each part can be performed using either the coordinate- 
or the momentum-space expression,
depending on the convenience. For instance, if the selected potential energy
model includes the CD Bonn 2000 potential and the (point) Coulomb interaction, 
the matrix elements of the CD Bonn 2000 are computed using 
Eq.~(\ref{eq:potq}), the ones of the Coulomb potential are computed using 
Eq.~(\ref{eq:potrloc}). The same procedure can be applied to the 
TNIs mostly used in the literature, i.e. the Urbana-type~\cite{Pud95} and 
Tucson-Melbourne-type~\cite{Coo79,Rob86} potentials.
However, since the aim of this work is to study 
the applicability of the HH method when non-local potentials are used, 
the TNI has not been included.

Other potentials models which are frequently used in the literature are 
those developed by Doleschall {\it et al.}~\cite{DB00,Dea03}. These models
consist of non-local operators given in coordinate space. In that case, 
it is more convenient to perform the calculation using 
Eq.~(\ref{eq:potr}).

The calculation of the integrals involved in 
Eqs.~(\ref{eq:potr}) or~(\ref{eq:potq})
is a non-trivial numerical task. 
As an example, let us consider  Eq.~(\ref{eq:potq}). 
Remembering that $\widetilde V(\bfk'_N;\bfk_N)$ acts on the particle pair
$1$,$2$, it is convenient, using the Raynal-Revai coefficients and the recoupling of
spin-isospin states, to express the states $\langle\Psi_{ \{G'\}, l' }|$ and
$|\Psi_{  \{G\}, l}\rangle$ in momentum space as a superposition of HH
functions  and spin-isospin states constructed using the order of particles
$1,2,3,\ldots,A$. Most of the integrations can be now performed analytically
and the matrix elements, for a general $A$-body system, reduces to a sum of
three-dimensional integrals of the type
\begin{eqnarray}
   I^{l',l;j,S',S}_{G',\ell_N',n_N';G,\ell_N,n_N} &=&
   \int_0^\infty dq\; q^{D-4} \int_0^\infty dk_N\; (k_N)^2 \int_0^\infty
    dk_N' (k_N')^2 g_{G',l'}(Q') \nonumber \\    
  &&\qquad \times  (\cos\varphi_N')^{\ell_N'} (\sin\varphi_N')^{\nu'}
   P_{n_N'}^{\nu'+D/2-5/2,\ell_N'+1/2}(\cos2\varphi_N')
   v^j_{\ell_N',S';\ell_N,S}(k_N',k_N) \nonumber \\
   && \qquad \times  g_{G,l}(Q)  (\cos\varphi_N)^{\ell_N} (\sin\varphi_N)^{\nu}
   P_{n_N}^{\nu+D/2-5/2,\ell_N+1/2}(\cos2\varphi_N)
   \delta_{\nu,\nu'}\ , \label{eq:iii}
\end{eqnarray}
where 
\begin{equation}
  Q^2=k_N^2+q^2\ ,\quad (Q')^2=(k_N')^2+q^2\ , \quad\cos\varphi_N=k_N/Q
  \ , \quad \cos\varphi_N'=k_N'/Q'\ ,
\end{equation}
and 
\begin{equation}
  \nu=G-2n_N-\ell_N\ , \qquad \nu'=G'-2n_N'-\ell_N' \ .
\end{equation}
Moreover, in Eq.~(\ref{eq:iii})
$v^j_{\ell_N',S';\ell_N,S}(k_N',k_N)$ is the two-body
potential acting between   states of the pair of particles $1,2$
of total angular momentum $j$, and orbital angular momentum  and spin quantum
numbers $\ell_N',S'$ (on the left) and $\ell_N,S$ (on the right). 
The integrals $I^{l',l;j,S',S}_{G',\ell_N',n_N';G,\ell_N,n_N}$ can be computed
beforehand and stored in computer disks. The last step consists in combing
$I^{l',l;j,S',S}_{G',\ell_N',n_N';G,\ell_N,n_N}$ with the Raynal-Revai
coefficients to obtain the matrix elements of Eq.~(\ref{eq:potq}).
Finally, the integrations involved in Eq.~(\ref{eq:iii})
can be accurately performed with standard numerical techniques
(Gauss integration)~\cite{Abr72}.

\section{Results}
\label{sec:res}

In this section, the binding energy and ground-state properties 
obtained for the nuclear systems with $A$=3 and 4 are presented for 
the CD Bonn 2000~\cite{Mac01}
and the N3LO~\cite{EM03} momentum space potentials. 
These interaction models are decomposed on partial waves, and 
the partial wave decomposition 
is truncated at a certain value of the two-body total angular momentum 
$j_{max}$. In the present work $j_{max}$=6 has been chosen, 
which allows for an 
accuracy of better than 1 keV in the triton binding energy. 

The section  is divided into three subsections: in 
Sec.~\ref{subsec:conva3} and~\ref{subsec:conva4} the convergence 
of the expansion with respect to the quantum numbers $\{G\}$ and the 
number $M$ of hyperradial functions is discussed 
for $A$=3 and 4, respectively. The converged results for the 
ground-state properties of triton, $^3$He and $^4$He are presented 
in Sec.~\ref{subsec:resa34}, and compared with the results 
obtained with different approaches.

\subsection{Convergence of the $A$=3 Results}
\label{subsec:conva3}

In this subsection, we 
show the level of accuracy reached by the method presented in 
this work. As an example, in Table~\ref{tab:conve} and~\ref{tab:convl}, 
the triton binding energy 
is calculated with the CD Bonn 2000 potential 
using only the first 3 channels of Table I of Ref.~\cite{Kie97a},
and increasing the value of the grand angular momentum $G$ and the value of 
the number $M$ of basis elements in the expansion of the 
hyperradial and hypermomentum functions $f_l(\rho)$ and 
$g_{G,l}(Q)$ (see Eqs.~(\ref{eq:rexp}) and~(\ref{eq:qexp})).
In Table~\ref{tab:conve} the exponential basis of Eqs.~(\ref{eq:flexp})
and~(\ref{eq:glexp}) is used, 
and the non-linear parameters $\beta$ and 
$\gamma$ are 0.78 fm$^{-1}$ and 1.2 fm$^{-1}$  respectively. 
In Table~\ref{tab:convl},  
$f_l(\rho)$ and $g_{G,l}(Q)$ are expanded on 
Laguerre polynomials, Eqs.~(\ref{eq:fllag}) and~(\ref{eq:gllag}), 
and the non-linear parameter $\gamma$ is 4.0 fm$^{-1}$.
 
By inspection of the tables, we can conclude that 
i) the convergence of the exponential basis at the 1 keV accuracy level 
is slightly faster than the one 
of the Laguerre polynomials basis. In fact, the maximum value of 
$M$ used in the two expansions, $M_{max}$, is 16 and 24 respectively. 
However, it should be noted that 
the exponential basis gives numerical problems for higher values 
of $M_{max}$, and turns out to be not suitable for the N3LO 
potential. Nevertheless, the binding energy $B$ at $G$=80 calculated with 
$M_{max}=16$ exponentials is only 0.5 keV smaller than 
the result obtained with $M_{max}=24$ Laguerre polynomials.
ii) In the case of the exponential basis, 
the expansion on $M$ is truncated when 
$B(M_{max})-B(M_{max}-2)<$ 0.5 keV. In the case of the 
Laguerre polynomial expansion, $B(M_{max})-B(M_{max}-2)$ is smaller 
than a tenth of keV. 
iii) The convergence on $G$ is reached at $G_{max}$=80. In fact, in 
Table~\ref{tab:conve}, we can see that going from 
$G$=60 to $G$=70 and from $G$=70 to $G$=80, the gain in $B$ is 
1.3 keV and 0.5 keV, respectively. In Table~\ref{tab:convl}, the 
corresponding values are 1.3 keV and 0.6 keV, respectively.
Due to these considerations, we are confident in 
concluding that the results of 7.699 MeV 
for the triton binding energy, obtained with only the first 
3 channels in our channel expansion, is accurate at least 
at the 1 keV level. Note that the value for $G_{max}$ used 
in Ref.~\cite{Kie97a} for the AV18 potential and the first 3 
expansion channels is 180. This is related to the fact that both the 
CD Bonn 2000 and N3LO potential models have a rather 
soft short-range repulsion compared with the AV18 potential model. 

A similar procedure to reach the convergence on $G$ and $M$ has 
been used also when other channels are included in the calculation. 
The converged $A$=3 results presented in Sec.~\ref{subsec:resa34}
have been obtained including up to 23 channels, 
18 having total isospin $T$=1/2 
and 5 having $T$=3/2 (see Sec.~\ref{sec:cal} and Ref.~\cite{Kie97a}). 
In particular, for the CD Bonn 2000 (N3LO) potential, 
the maximum value of the grand angular momentum $G_{max}$ is 
80 (40) for the first 3 channels, 40 (20) for the next 5 ones, 
and 30 (20) for the last 10 channels with $T$=1/2. The convergence 
for the $T$=3/2 channels is reached using $G_{max}$=20. 

\begin{table}
\caption[Table]{Triton binding energies in MeV, calculated with the 
CD Bonn 2000 two-nucleon interaction, using only the first 3 
expansion channels and the exponential functions 
of Eqs.~(\protect{\ref{eq:flexp}}) and~(\protect{\ref{eq:glexp}})
as expansion basis for the hyperradial and hypermomentum functions.
$M$ is the maximum number of basis elements, $G$ is the 
grand angular momentum.}
\begin{tabular}{c@{$\quad$}l@{$\quad$}l@{$\quad$}l@{$\quad$}l@{$\quad$}l}
\hline
$G$ & $M$=8 & $M$=10 & $M$=12 & $M$=14 & $M$=16 \\ 
\hline
40 & 7.6764 & 7.6835 & 7.6842 & 7.6845 & 7.6846 \\
50 & 7.6845 & 7.6921 & 7.6934 & 7.6934 & 7.6935 \\
60 & 7.6870 & 7.6948 & 7.6958 & 7.6962 & 7.6963 \\
70 & 7.6880 & 7.6958 & 7.6969 & 7.6974 & 7.6976 \\
80 & 7.6885 & 7.6964 & 7.6974 & 7.6979 & 7.6981 \\
\hline
\end{tabular}
\label{tab:conve}
\end{table}

\begin{table}
\caption[Table]{Same as Table~\protect\ref{tab:conve}
but with the Laguerre polynomial 
expansion of Eqs.~(\protect{\ref{eq:fllag}}) and~(\protect{\ref{eq:gllag}})
for the hyperradial and hypermomentum functions.}
\begin{tabular}{c@{$\ $}|@{$\ $}l@{$\ $}l@{$\ $}l@{$\ $}l@{$\ $}l@{$\ $}
                l@{$\ $}l@{$\ $}l@{$\ $}l}
\hline
$G$ & $M$=8 & $M$=10 & $M$=12 & $M$=14 & $M$=16 & $M$=18 & $M$=20
    & $M$=22 & $M$=24 \\ 
\hline
40 & 7.6582 & 7.6773 & 7.6827 & 7.6835 & 7.6842 & 7.6846 & 7.6847 & 7.6848 
& 7.6848 \\
50 & 7.6665 & 7.6862 & 7.6916 & 7.6924 & 7.6931 & 7.6936 & 7.6937 & 7.6937 
& 7.6937 \\
60 & 7.6691 & 7.6892 & 7.6946 & 7.6955 & 7.6962 & 7.6966 & 7.6967 & 7.6967 
& 7.6967 \\
70 & 7.6702 & 7.6904 & 7.6959 & 7.6967 & 7.6974 & 7.6978 & 7.6979 & 7.6980 
& 7.6980 \\
80 & 7.6706 & 7.6910 & 7.6965 & 7.6973 & 7.6980 & 7.6984 & 7.6985 & 7.6986 
& 7.6986 \\
\hline
\end{tabular}
\label{tab:convl}
\end{table}

\subsection{Convergence of the $A$=4 Results}
\label{subsec:conva4}

First of all, we discuss the convergence of the expansion with respect to
the quantum numbers $\{G\}$. We follow the procedure adopted in
Ref.~\cite{Viv05}. The states $| \Psi_{ \{G\},l}\rangle$ are separated in six
{\it classes}, depending on the
total isospin quantum number $T$ and 
on $\ell_1,\ell_2,\ell_3$, the values of the orbital angular momentum
quantum numbers associated to the three Jacobi vectors (or momenta). 
The most important classes are those ones 
with $T=0$ and with the lowest values of $\ell_1+\ell_2+\ell_3$. 
The convergence of the class $i$ is studied by including in the expansion all
the states $|\, \Psi_{ \{G\}, l}\, \rangle$   with grand angular quantum
number $G\le G_i$, and then
increasing the value of $G_i$. The rate of convergence depends primarily on
the repulsion at short interparticle distance of the adopted
potential models. More details on this argument 
and the choice of the basis can be found in Ref.~\cite{Viv05}.

In the case of the two potential models considered in this work, 
it turns out that both the models have a rather soft repulsion and
the convergence is reached for values of $G_i$, $i=1,\ldots,6$ rather smaller
than those found in Ref.~\cite{Viv05}, 
where the the AV18~\cite{AV18} and Nijmegen II~\cite{Nijm} 
potential models were considered. 
To give an example,  for those models the
convergence was reached for 
$\{G_1,G_2,G_3,G_4,G_5,G_6\}=\{72,40,34,28,24,20\}$, while
for the CD Bonn 2000 potential a satisfactory convergence is obtained for
$\{G_1,G_2,G_3,G_4,G_5,G_6\}=\{52,34,28,24,20,20\}$. The N3LO potential 
is derived from an effective field theory by means of an expansion valid at low
momenta, and it is strongly suppressed at high momenta. This corresponds in
coordinate space to a rather soft repulsion at short interparticle
distances. Correspondingly, a satisfactory convergence is reached at 
$\{G_1,G_2,G_3,G_4,G_5,G_6\}=\{24,20,16,16,16,20\}$. The ``missing binding
energy'' due to the truncation of the basis has been estimated, 
using the same procedure adopted in Ref.~\cite{Viv05}, to be around
$10$ keV for the CD Bonn 2000 case, and around $5$ keV for the N3LO
potential.

We now consider the convergence with respect to the number $M$ of
functions $g_{G,l}(Q)$ used in Eq.~(\ref{eq:wafu1}). 
In Table~\ref{tab:conv4he}, the binding energy and other ground state
properties calculated for the N3LO potential for different values of $M$
are reported. In this example, the hyperradial functions have been 
expanded on Laguerre polynomials with $\gamma=6.0$ fm$^{-1}$. 
All the HH functions belonging to the six classes discussed above, 
namely for the
choice $\{G_1,G_2,G_3,G_4,G_5,G_6\}=\{24,20,16,16,16,20\}$, 
have been included.
From the table it is seen that all quantities have converged quite 
well for $M\approx16-18$.

\begin{table}
\caption[Table]{\label{tab:conv4he}
Convergence of $\alpha$--particle binding energy $B$ (MeV), mean value of the
kinetic energy $\langle T\rangle$ (MeV), mean square radius $\sqrt{\langle r
\rangle^2}$ (fm) and $P-$ and $D-$wave percentages for different values of
$M$, the number of functions $g_{G,l}(Q)$ included in the expansion. In this
example, $g_{G,l}(Q)$ are obtained expanding the hyperradial functions
on Laguerre polynomials with $\gamma=6.0$ fm$^{-1}$. The potential model
considered is the N3LO. The basis includes 
HH states with $\{G_1,G_2,G_3,G_4,G_5,G_6\}=\{24,20,16,16,16,20\}$ (see the
text for more details).}
\begin{tabular}{c@{$\quad$}c@{$\quad$}c@{$\quad$}c@{$\quad$}
                c@{$\quad$}c@{$\ $}}
\hline
$M$ & $B$ & $\langle T\rangle$ 
& $\sqrt{\langle r^2 \rangle}$ & $P_P$ & $P_D$ \\
\hline
  8 &  24.649 &  72.326 & 1.453 & 0.172 &  9.272 \\
 10 &  25.195 &  69.940 & 1.495 & 0.173 &  9.315 \\
 12 &  25.339 &  69.309 & 1.512 & 0.172 &  9.287 \\
 14 &  25.362 &  69.244 & 1.514 & 0.172 &  9.290 \\
 16 &  25.373 &  69.230 & 1.516 & 0.172 &  9.289 \\
 18 &  25.376 &  69.236 & 1.516 & 0.172 &  9.289 \\
\hline
\end{tabular}
\end{table}

\subsection{Results for $A$=3 and 4}
\label{subsec:resa34}

The converged 
values for the triton, $^3$He and $^4$He binding energies and 
other ground-state properties are listed in
Table~\ref{tab:3h},~\ref{tab:3he} and~\ref{tab:4he}, respectively.
All these results have been obtained 
expanding the hyperradial functions on Laguerre polynomials, 
with $\gamma$=4 fm$^{-1}$ and 7 fm$^{-1}$ for the
CD Bonn 2000 and N3LO potentials, respectively. 
The numerical uncertainty in the 
triton and $^3$He binding energy has been estimated to be at
the most of the order of 1 keV, and in the $^4$He binding energy of the order
of 10 keV. These results are compared with those 
obtained with other approaches.
In particular, we have considered the 
Faddeev (for $A$=3) and Faddeev-Yakubovsky (FY) (for $A$=4) 
approach~\cite{Nog03,Nogpv,Del03,Delpv},
and the No Core Shell Model (NCSM) approach~\cite{Nav04,Navpv}.
For sake of comparison, 
the results for the AV18 potential are also 
listed~\cite{Kie97a,Viv05,Nog03,Nea02,LC04}.
Note that the HH results of Ref.~\cite{Kie97a} do not include the $T=3/2$ states.
In the $^3$He and $^4$He case, all the results 
for the CD Bonn 2000 and N3LO potentials have been obtained including 
the point Coulomb interaction, except for the $^4$He FY results with the
N3LO potential. The latter have been obtained including a
more complicated electro-magnetic interaction between nucleons~\cite{Nogpv},
whose effects are however quite similar to those of the point Coulomb
interaction. 

There is a good agreement, at the level of 0.1 \%, 
among the results obtained with different 
techniques for all quantities considered in the tables. 
Furthermore, the mean value
of the kinetic energy for the N3LO case is smaller than that found
with the AV18 potential. This is due to the fact that
the repulsion at short interparticle distances is softer for the N3LO
potential than for the AV18 potential. Also the
percentages of the $P$- and $D$-waves is significantly smaller for the N3LO
potential. This explains the faster convergence 
of the HH expansion in this
case. The CD Bonn 2000 is intermediate between the two cases discussed above.

We now consider the results for the $T$=3/2 and $T> 0$ components 
in the $A$=3 and 4 systems, respectively.
The percentage of the $T=1$ component 
in the $^4$He ground state wave function is almost independent from
the adopted potential model. In fact, as discussed in Ref.~\cite{Viv05},
50\%  of it is due to the effect of the Coulomb interaction between
the protons, the remaining 50\% is due to the charge symmetry breaking 
(CSB) terms
in the nuclear interaction. Consequently, the difference in $P_{T=1}$ for 
the various two-body potentials is small. 
On the contrary, the $T$=3/2 component in triton and the 
$T=2$ component in $^4$He are largely dominated by CSB of nuclear
origin (different pion masses, etc.). The values reported in
Table~\ref{tab:3h} and~\ref{tab:4he} show that, depending on the interaction, 
rather  different values for the triton $P_{T=3/2}$ and 
the $^4$He $P_{T=2}$ 
are obtained (note that the standard models of TNI have
little effect on the isospin admixtures~\cite{Viv05}). The CSB magnitude
in the two-body interaction is fixed by fitting the $NN$ scattering
data, and therefore comes from differences observed
in the $pp$ and $np$ systems. In particular, the CD Bonn 2000 and N3LO
potentials fit the same $NN$ data set~\cite{Maclpv}. The origin of the rather
large differences found for the triton $P_{T=3/2}$ and the $^4$He 
$P_{T=2}$ (a factor 5 between CD Bonn 2000 and
N3LO) must be related to quite different off-shell behavior of
the CSB terms of both interactions. Note that the 
$T$=3/2 component in $^3$He is also affected by the 
Coulomb interaction, which reduces by more than a factor of 2 
the difference between the CD Bonn 2000 and N3LO $P_{T=3/2}$ results.

Finally, we observe that knowledge of the $T=1$ and $2$ 
percentages could be 
important for parity-violating electron scattering experiments on ${}^4$He,
aimed at studying admixture of strange quark $s\bar s$ pairs
in nucleons and nuclei~\cite{BmK01,Happex,RHD94}.
It could play an important role also in the study of the
reaction $d+d\rightarrow \alpha+\pi^0$. This reaction is possible only if 
isospin symmetry is violated, namely it probes directly the CSB terms in the
nuclear Hamiltonian~\cite{IUCF2,Gea04}.

\begin{table}
\caption[Table]{\label{tab:3h}
The triton binding energies $B$ (MeV), the mean 
square radii $\sqrt{\langle r^2\rangle}$ (fm), 
the expectation values of the kinetic energy operator $\langle T\rangle$
(MeV), and the mixed-symmetry $S'$, $P$, $D$, and 
isospin $T$=3/2 probabilities (all in \%), calculated with the 
CD Bonn 2000 and N3LO potentials, are compared 
with the results obtained within the 
Faddeev equations approach~\protect{\cite{Nog03,Nogpv,Del03,Delpv}} (FE)
and within the No Core Shell Model 
(NCSM) approach~\protect{\cite{Nav04,Navpv}}. 
The results obtained in Refs.~\protect\cite{Kie97a}
and~\protect\cite{Nog03} within the HH, 
correlated-hyperspherical-harmonics (CHH) and FE approaches 
for the AV18 potential have been also 
reported for sake of comparison. These last HH results do 
not include the $T$=3/2 states.}
\begin{tabular}{l@{$\quad$}l@{$\qquad$}c@{$\quad$}
                c@{$\quad$}c@{$\quad$}c@{$\quad$}c@{$\quad$}c@{$\quad$}c}
\hline
Interaction   & Method & $B$ & 
              $\langle T\rangle$  &
              $\sqrt{\langle r^2\rangle}$ &
              $P_{S'}$ & $P_P$ & $P_D$ & $P_{T=3/2}$ \\
\hline
CD Bonn 2000
    & HH~(this work)     & 7.998 & 37.630 &
      1.721 & 1.31 & 0.047 & 7.02 & 0.0049 \\
    & FE~\protect\cite{Nog03,Nogpv}  & 7.997 & 37.620 &
      -  & 1.31 & 0.047 & 7.02 & 0.0048 \\
    & FE~\protect\cite{Del03,Delpv}  & 7.998 & 37.627 &
      -  & 1.31 & 0.047 & 7.02 & 0.0048 \\
    & NCSM~\protect\cite{Nav04,Navpv}  & 7.99(1) & -  &
      - & - & -  & - & - \\
\hline
N3LO
    & HH~(this work)     & 7.854 & 34.555 &
      1.758 & 1.36 & 0.037 & 6.31 & 0.0009 \\
    & FE~\protect\cite{Nog03,Nogpv}  & 7.854 & 34.546 &
      -  & 1.37 & 0.037 & 6.32 & 0.0009 \\
    & FE~\protect\cite{Del03,Delpv}  & 7.854 & 34.547 &
      -  & 1.37 & 0.037 & 6.32 & 0.0009 \\
    & NCSM~\protect\cite{Nav04,Navpv}  & 7.85(1) & -  &
      - & - & -  & - & - \\
\hline
AV18
    & CHH~\protect\cite{Nog03}  & 7.624 & 46.727 &
       -    & 1.293 & 0.066 & 8.510 &  0.0025  \\
    & HH~\protect\cite{Kie97a}  & 7.618 & 46.707 &
      1.770 & 1.294 & 0.066 & 8.511 &    -     \\
    & FE~\protect\cite{Nog03}  & 7.621 & 46.73  &
      -  & 1.291 & 0.066 & 8.510 & 0.0025 \\
\hline
\end{tabular}
\end{table}

\begin{table}
\caption[Table]{\label{tab:3he}
Same as Table~\protect\ref{tab:3h} but for $^3$He.}
\begin{tabular}{l@{$\quad$}l@{$\qquad$}c@{$\quad$}
                c@{$\quad$}c@{$\quad$}c@{$\quad$}c@{$\quad$}c@{$\quad$}c}
\hline
Interaction   & Method & $B$ & 
              $\langle T\rangle$  &
              $\sqrt{\langle r^2\rangle}$ &
              $P_{S'}$ & $P_P$ & $P_D$ & $P_{T=3/2}$ \\
\hline
CD Bonn 2000
    & HH~(this work)     & 7.262 & 36.777 &
      1.759 & 1.54 & 0.046 & 7.00 & 0.0109 \\
    & FE~\protect\cite{Nog03,Nogpv}  & 7.261  & 36.756 &
      -  & 1.54 & 0.046 & 7.00 & 0.0110 \\
    & FE~\protect\cite{Del03,Delpv}  &  7.263 & 36.761 &
      -  & 1.54 & 0.046 & 7.00 & 0.0110 \\
\hline
N3LO
    & HH~(this work)     & 7.128 & 33.789 &
      1.797 & 1.61 & 0.037 & 6.31 & 0.0062 \\
    & FE~\protect\cite{Nog03,Nogpv}  & 7.128 & 33.775 &
      -  & 1.61 & 0.037 & 6.31 & 0.0063 \\
    & FE~\protect\cite{Del03,Delpv}  & 7.128 & 33.775 &
      -  & 1.61 & 0.037 & 6.32 & 0.0063 \\
\hline
AV18
    & CHH~\protect\cite{Nog03}  & 6.925 & 45.685 &
       -    & 1.530 & 0.065 & 8.467 &  0.0080  \\
    & FE~\protect\cite{Nog03}  & 6.923 & 45.68  &
      -  & 1.524 & 0.065 & 8.466  & 0.0081 \\
\hline
\end{tabular}
\end{table}

\begin{table}
\caption[Table]{\label{tab:4he}
The $\alpha$--particle binding energies $B$ (MeV), the mean 
square radii $\sqrt{\langle r^2\rangle}$ (fm), 
the expectation values of the kinetic energy operator $\langle T\rangle$
(MeV), and the $P$, $D$, $T=1$ and $T=2$ probabilities (\%) for the two
non-local potentials considered in this paper. The results obtained
in Ref.~\protect\cite{Viv05} for the AV18 potential have been also reported for
sake of comparison. The  results obtained by other techniques 
are also listed.}
\begin{tabular}{l@{$\quad$}l@{$\qquad$}c@{$\quad$}
                c@{$\quad$}c@{$\quad$}c@{$\quad$}c@{$\quad$}c@{$\quad$}c}
\hline
Interaction   & Method & $B$ & 
              $\langle T\rangle$  &
              $\sqrt{\langle r^2\rangle}$ &
              $P_P$ & $P_D$ & $P_{T=1}$ & $P_{T=2}$ \\
\hline
CD Bonn 2000
    & HH~(this work)     & 26.13 & 77.58 &
      1.454 & 0.223 & 10.74 & 0.0029 & 0.0108 \\
    & FY~\protect\cite{Nogpv}  & 26.16    & 77.59  &
      -  & 0.225 & 10.77 & 0.0030 & 0.0108 \\
\hline
N3LO
    & HH~(this work)           & 25.38 & 69.24 &
      1.516 & 0.172  & 9.289 & 0.0035 & 0.0024  \\
    & FY~\protect\cite{Nogpv}  & 25.37 & 69.20  &
      - & 0.172 & 9.293  & 0.0033 &  0.0024 \\
    & NCSM~\protect\cite{Nav04}  & 25.36(4) & -  &
      - & - & -  & - & - \\
\hline
AV18
    & HH~\protect\cite{Viv05}  & 24.210 &  97.84 & 
      1.512  & 0.347  & 13.74 & 0.0028 & 0.0052 \\
    & FY~\protect\cite{Nea02}  & 24.25  &  97.80 & 
       -     & 0.35   & 13.78 & 0.003 & 0.005 \\
    & FY~\protect\cite{LC04}   & 24.223   & 97.77 & 
      1.516  &  -  &  -  &  -  & -  \\
\hline
\end{tabular}
\end{table}

\section{Summary and Conclusions}
\label{sec:con}

The $A=3$ and 4 nuclear ground states have been studied with
non-local two-body potentials using the HH method.
The variational wave function is written as an expansion over a complete basis,
which is constructed in coordinate space as a product of hyperradial,
HH and spin-isospin functions. The main task is 
the calculation of the Fourier transform of the expansion basis. 
However, given the
properties of the HH functions, this Fourier transform reduces to a
one-dimensional integral which can
be obtained analytically. The application is therefore rather straightforward,
and the matrix elements of a given two-body 
interaction can always be reduced to
three-dimensional integrals (two-dimensional for local potentials).
We have presented  the results for the CD Bonn 2000 and N3LO  $NN$
potential models. In both cases we have found very good agreement with the
results obtained by other groups. We have also given the estimates obtained for
various ground state properties and pointed out that the different models
predict very different (up to a factor 5) percentages of the isospin
admixtures in the ${}^3$H and ${}^4$He ground states. 

In recent years, new models of nuclear interaction have been 
constructed from chiral perturbation theory. 
They are non-local and given in momentum
space~\cite{EGM00}. The N3LO potential is the first potential of this 
type fitting the $NN$ data set with $\chi^2$/datum $\simeq$ 1. 
Another class of recently developed
potentials are the so-called {\it low-q} potentials~\cite{BKS03,Cor05}. They
are ``renormalized'' two-body potentials $\widetilde V_{NN}(k,k')$,
given in momentum space, where the tail for high values of $k$ and $k'$
has been eliminated. 

It is important, therefore, to have accurate techniques to solve
the Schr\"odinger equation for non-local momentum-space potentials. Up to
now, only the FY~\cite{Nea02,Del03} and the NCSM~\cite{Nav04} methods were
available for this task. In this paper, we have shown that also the HH method
can be successfully applied to treat this kind of potentials. 

There are two other important motivations behind this work. The first
one is that the HH formalism can be applied also to
scattering problems. As discussed in Sec.~\ref{sec:intro}, 
there are still a number of theoretical 
problems in the $N-d$ and $p-{}^3$He reactions, 
and it would be very interesting
to check whether the use of these new potential models could solve
these problems. In particular, 
it appears promising the development of more
realistic models of TNIs to be tested in $A=3$ and $4$ nuclear 
systems. The second motivation is the possibility of the
extending the HH method to larger systems. The feasibility of such an
application would require the solution of several problems, 
having to do with the fast
computation of the Raynal-Revai coefficients for $A>4$ and the very large
degeneracy of the basis. However, since these new potentials are softer than
the older ones, it should be not too difficult to solve this latter problem. 
Work in both directions is currently underway.

\section*{Acknowledgment}
The authors would like to thank A.\ Deltuva, A.\ Nogga and 
P.\ Navr\'atil for providing their Faddeev and No Core Shell Model results
and R. Machleidt for useful discussions.
The authors would like to thank 
A.\ Nogga also for pointing them out that in the 
CD Bonn 2000 and N3LO potentials codes a minus sign is omitted in the 
coupled channels. Finally, the authors would like to thank R.\ Schiavilla 
for a critical reading of the manuscript.


\end{document}